\documentclass[aps,twocolumn,showkeys]{revtex4}

\pdfoutput=1
                                
\usepackage{graphicx}                                       
\usepackage{amssymb}
\usepackage{wasysym}

\newcommand{\umafigura}[3]{   
  \begin{figure}[!t]
    \centering
    \includegraphics[width=2.5in]{#1}
    \caption{#3}
    \label{#2}
  \end{figure}
}

\newcommand{\umafigural}[3]{   
  \begin{figure}[!t]
    \centering
    \includegraphics[width=\linewidth]{#1}
    \caption{#3}
    \label{#2}
  \end{figure}
}

\newcommand{\eqt}[2]{
    \begin{equation}
             {#2}
             \label{#1}
    \end{equation}
}

\begin{document}
%
%
%

\title{Non-equilibrium Thermodynamics of a Resistance Element}


\author{
  F. A. Silveira${}^{}$\footnote{Email address: {\it fsilveira@inmetro.gov.br} (corresponding author)} and
  R. T. B. Vasconcellos${}^{}$\footnote{Email address: {\it rtbvasconcellos@inmetro.gov.br}}
}

\affiliation{
  Instituto Nacional de Metrologia, Qualidade e Tecnologia,\\
  Avenida N. S. das Gra\c cas 50, 25250-020 D. Caxias RJ, Brazil}

\date{\today}


\begin{abstract}
   This paper describes the thermal characterization of a coaxial
   calculable resistor, part of traceability chain of the capacitance
   unit to the quantum Hall effect, in construction at Inmetro. The
   realization of the capacitance unit (farad) is related to the QHE by
   a calculable coaxial resistor, and three low-uncertainty coaxial
   bridges: the two-terminal and the four-terminal bridges, already in
   operation at Inmetro; and the quadrature bridge, in final stage of
   construction.
\end{abstract}


\maketitle

%
%
%
\section{Introduction}
\label{intro}

One of the mid-term projects of the 
Laboratory of Metrology of Electrical Standards at Inmetro
(Lampe) is to establish the traceability of the capacitance unit all
the way to the quantum Hall effect (QHE) \cite{vasconcellos2009}.
The traceability chain from the farad to the quantum Hall resistance
(QHR) begins with the AC-DC transference of the resistance unit
through a calculable resistor. The next steps consist in comparisons
in high-precision coaxial in-phase and quadrature bridges
\cite{kibble,gibbings1963,vasconcellos2010}.

Volatility in the value of this calculable resistor can be caused by the
degradation of the resistive element surface on welds, which brings
as an immediate effect the increase of the contact resistance at
junction points and very likely a reduction of the repetitivity of
the measurements. Furthermore, if the element finds itself under
mechanical strain, there can be a slow migration of macroscopic
physical properties of the resistor, which may cause long-term drift
\cite{kucera2009}.

We propose the introduction of a electrically inert support
for the resistive element -- usually of difficult handling due to its
reduced dimensions -- in order to allow its better settlement to the
internal terminals in soldering, as well as a finer control of its nominal
value. To this end, a resistor prototype was built, and some of its
physical properties were studied in this work.

In the sections that follow, some brief comments are made on general thermal properties
of metals. Details on the resistor element and its support construction
are presented in Sec. \ref{construcao}. Still in that section, we present
a physical characterization of the resistive element, which can be useful for
future reference, and some conclusions are left to Sec. \ref{conclusao}.

\section{Construction}
\label{construcao}

In the design adopted by Lampe, the resistor is made of a nickel,
chrome, aluminum and silicon alloy, named {\it isaohm}. As a rule, the
resistor construction must be guided by the symmetries of the
electromagnetic fields induced by the resistive element. The trivial
symmetry, which gives rise to the most simple construction, is the
cylindrical one \cite{gibbings1963,jackson}. The coaxial resistor made
at Inmetro follows the 4-connector, 2-plane design described in
Refs. \cite{trapon2003,delahaye2005}. The resistor element consists
of a isaohm wire of $\diameter$0.014 mm, approximately
11 cm long. A few physical properties of isaohm are presented in
Ref. \cite{isaohmtable}.

A sheet of epoxy laminate of width 1.60 mm and plane
dimensions 1.030$\times$16.700 cm was cut in order to host two copper islands
11.080 cm apart. The two ends of the isaohm wire that make the resistive
element are then locked to the islands, together with ordinary
$\diameter$1 mm copper wires to allow resistance measurements in
the 4-terminal scheme.

The resistor so constructed has nominal value 1080.02 $\Omega$, and final linear resistivity
about 8.5\% above the result of measurements performed before the wire soldering process.
This set makes the test prototype for the
calculable resistor to be used in the derivation of the AC ohm
from QHR measurements \cite{fsilveira2012}.

%
\subsection{Thermal Inertia and Characteristic Time}
\label{efeitostermoeletricos}

The resistive element described in Sec. \ref{construcao} has thermal
inertia and capacity which depend on equilibrium thermodynamic
properties, such as specific heats, and bulk and surface transport
properties, such as the electrical and thermal conductivities, and
heat transference rates. If the frequency $f$ of the function
applied to the resistor (voltage or current) is such that
$f^{-1}<<\tau$ (where $\tau$ is the thermal stabilization time of
the resistor), the thermal oscillations induced by thermoelectric
effects \cite{macdonald} become irrelevant in practice.

\umafigural{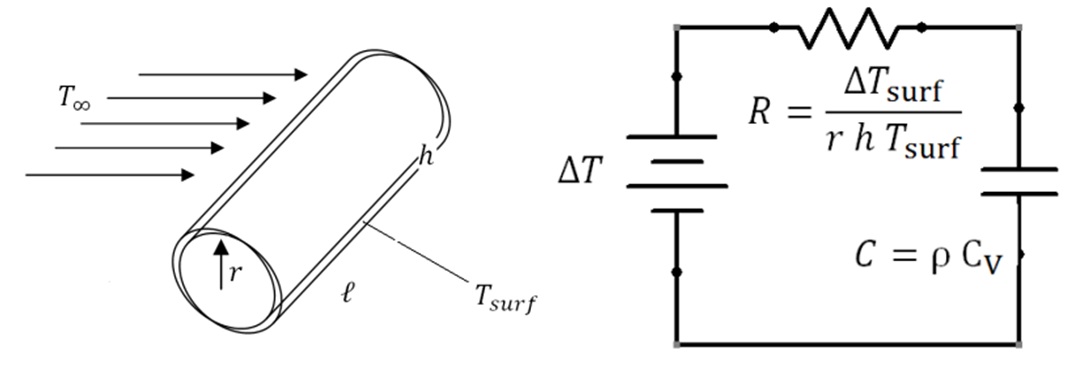}{lumpmodel}{The diagrams show how the lumped parameter 
model for the heat transfer through the wire surface is constructed. In the illustrations,
$T_\infty$ stands for the heat bath temperature, and $T_{Surf}$ for the wire
surface temperature.}

This makes the high thermal integration time a desirable property
of resistors employed in electrical standardization, at the same
time that $\tau$ becomes an important construction parameter. A
rough estimate of the order of magnitude of $\tau$ can be realized
through a lumped parameter model for the heat conduction in the
resistive element analogous to the RC electrical circuit (see Fig. 
\ref{lumpmodel}). According to this model \cite{sissom},
   \eqt{tauchar}{\tau\sim {\ell c_V\rho/h},}
where $\ell$ is the resistive element length, $c_V$ is the specific heat of the
isaohm alloy, $\rho$ its mass density and $h$ the convection heat transference rate
through its surface. To isaohm,
$c_V\approx 0,46$ $J\cdot K^{-1}\cdot g^{-1}$ at 20$^o$C;
$\rho\approx 8\cdot Kg\cdot m^{-3}$ \cite{isaohmtable}; and
$h\sim 0,10$ $J\cdot m^{-2}\cdot K^{-1}\cdot s^{-1}$ \cite{sissom},
and it follows that $\tau\sim 400 s$.  Under these conditions,
thermal oscillations are integrated for frequencies above
$\tau^{-1}\sim 2,5\cdot 10^{-3}$ Hz, while capacitance bridge
measurements are usually taken at $10^4$ rad/s (approximately $1.6$
kHz) \cite{johnson2008}.

%
\subsection{Thermal Characteristics}
\label{caracteristicastermicas}

An active temperature control setup allowed that measurements were
performed under periodic fluctuations of temperature. The system consists
basically of a periodic signal generator that commands the thermostat
of a chamber, in which the test resistor is contained. Four- terminal resistance 
measurements are then  made through a 8-digit multimeter (Fig. \ref{setup}).

\umafigural{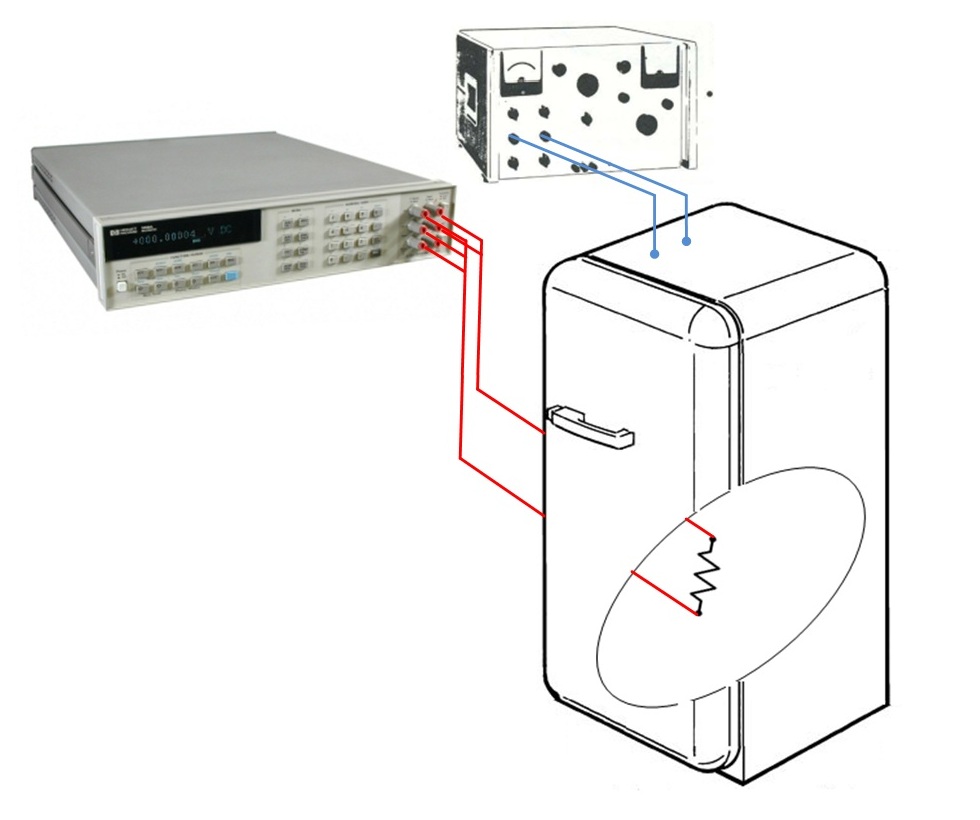}{setup}{Temperature control setup. The system showed
diagramatically in this illustration allowed us to make resistance measurements 
under periodic fluctuations of temperature.
}

This kind of measurement conditions, taken together with the lumped parameter
model adopted for heat conduction in the resistive element \cite{sissom},
enabled us to estimate, even if only roughly (because of the low resolution of
the environmental control thermostats), the characteristic time for thermal stabilization $\tau$,
defined in Sec. \ref{efeitostermoeletricos}. Still according to the model,
we expect that
\eqt{taufase}{\tau^{-1}=\omega \textrm{tan}^{-1}\left(\theta\right),}
where $\theta$ is the phase difference between the temperature and resistance
measurements made on the resistor; and $\omega$ is the oscilation frequency,
given in $rad\cdot s^{-1}$.

Figure \ref{fluttemp} shows how does normalized electrical resistance $R$ and
temperature $T$ vary with time $t$; time, showed on the horizontal axis, is
given in 2-second units. The temperature is measured with the internal thermometer
of the digital multimeter that makes the resistance measurements, an Agilent
3458A digital voltmeter, and swings in phase with the external laboratory
temperature, as we verified from direct measurements. The mean external and
multimeter internal temperatures are $23.0\ ^oC$ and $34.7\ ^oC$, respectively.

From the phase difference measurements between the curves showed in
this figure, we obtain $\omega\approx 2,51\cdot 10^{-3}\ rad\cdot
s^{-1}$ and $\theta\approx 1,4\ rad$, which gives $\tau\approx 420\
s$. This estimate, made entirely from resistance and temperature
time evolution measurements on the test prototype, is in excellent
agreement with the estimate made on Sec.
\ref{efeitostermoeletricos}, entirely based on simple dimensional
analysis.

\umafigura{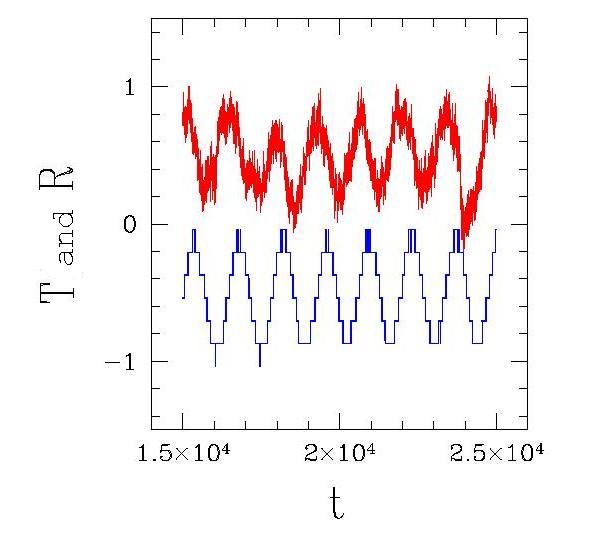}{fluttemp}{This figure shows how does normalized resistance $R$
(superior curve) and multimeter internal temperature $T$ as
functions of time $t$. From phase difference measurements on the
curves showed in this picture, we obtain $\tau\approx 420 s$, which
is in excellent agreement with the previous estimate made in Sec.
\ref{efeitostermoeletricos}.}

\section{Conclusions}
\label{conclusao}

At the early stages of the construction of a prototype calculable
resistor, we have run measurement tests under conditions of periodic
varying temperature. The resistor behavior when subjected to such
perturbation allowed the estimation of the characteristic time
$\tau$ for thermal stabilization from the time dispersion of
resistance measurements.

However, due to the cylindrical construction symmetry of the resistor, $\tau$
can also be independently estimated from thermodynamic (equilibrium) and
unsteady-state transport properties, as the heat transference rate or the
electrical and thermal conductivities, using a simple lumped parameter model,
such as the ones described in Ref. \cite{sissom}, for instance.

Both estimates of $\tau$ were compared, and good agreement was
found. This suggests the application of such models in the study and
characterization of calculable resistors such as Lampe's, with only
but very few restrictions, if any whatsoever. Among other things,
the estimation of $\tau$ can enable us to get a good grip of the
stabilization times of calibration systems, or aid us to correctly
quantify thermal properties of standard impedances in
non-steady-state regimes, usually of difficult characterization.



\end{document}